\documentclass[12pt]{article}
\usepackage[cp1251]{inputenc}
\usepackage{amsfonts}
\usepackage[english]{babel}
\usepackage{eufrak}
\usepackage{amssymb}

\begin{document}
\title{On Some New Proof of the Bogoliubov --- Parasiuk Theorem\\
(Nonequilibrium Renormalization Theory II )}
\sloppy
\author{D.V. Prokhorenko \footnote{Institute of Spectroscopy, RAS 142190 Moskow Region, Troitsk}}
\maketitle

\begin{abstract}
It is usually used a complicated combinatorics to prove the
Bogoliubov --- Parasiuk theorem. In the present paper we give a
proof of the Bogoliubov --- Parasiuk theorem which use a simple
combinatorics. To give this proof we interpret Feynman amplitudes
as distributions on the space of \(\alpha\)-parameters.

We will use this technique in the next paper to give a proof that
the divergences in nonequilibrium diagram technique can be
subtracted by means the counterterms of asymptotical state.
\end{abstract}
\newpage
\section{Introduction}
 \indent The mathematical theory of renormalization (\(R\)-operation) has been developed by
N.N. Bogoliubov and O.S. Parasiuk \cite{1,2}. K. Hepp has
elaborated on their proofs \cite{3}.

In standard framework based on Bogoliubov --- Parasiuk --- Hepp
sectors, renormalization theory needs in complicated
combinatorics. In the present paper we give a new proof of the
Bogoliubov --- Parasiuk theorem which needs simple combinatorics.
The main idea of this proof is an interpretation of the Feynman
amplitudes as distributions on the space of \(\alpha\) ---
parameters. The counterterms in our framework are local as in
\(\alpha\) --- parameters space as in coordinate space. This fact
makes possible to use decomposition of unite to solve the problem
of overlapping divergences.

Similar technique in renormalization theory have been used in \cite{4,5},
but in coordinate and momenta representations.

We will use the technique, developed in the present paper to prove
that the divergences in nonequilibrium diagram technique can be
renormalized by means of counterterms of asymptotical state.

\section{Definition of the Feynman Diagrams}
\textbf {Definition.} A Feynman graph is a triple
\newline
$$
 \Phi=(V,R,F),
$$
where \(V\) is a finite set, called the set of vertices,\newline
  \(R\) is a finite set, called the set of lines, and
  \newline
  \(f\) is a map: \newline
$$
 f: R\rightarrow(V\times V)\cup(V\times\{+,-\}).
$$

 The set \(R_{int}=f^{-1}(V\times V)\) is called the set of internal lines.
 The set \(R_{in}=f^{-1}(V\times \{+\})\) is called the set of
 lines coming into the graph. The set \(R_{out}=f^{-1}(V\times \{ - \})\) is called the set of lines
 coming from The Feynman graph. The set \(R_{ext}=R_{in}\cup R_{out}\) is called the set of external lines.
  \newline

 Let \(r\) be a line such that \(f(r)=(v_{1},v_{2})\). We say that the line \(r\) comes
 from the vertex \(v_{2}\) and comes into the vertex \(v_{1}\). We
 say also that the vertices \(v_{1}\) and \(v_{2}\) are connected
 by the line \(r\).

 Let \(r\) be a line such that \(f(r)=(v,+)\). We say that the line \(r\) is an external line coming
 into the vertex \(v\). Let \(r\) be a line such that \(f(r)=(v,-)\). In this case we say that the line
  \(r\) is an external line coming from \(v\).

 Below we will consider only graphs such that for each vertex \(v\) there exists a
 line \(r\) such that \(r\) comes into \(v\) or \(r\) comes from \(v\).

 By definition the path which connects the vertices \(v'\) and
 \(v''\) is a sequence of vertices
 $$
 v'=v_{0}, v_{1}, ... , v_{n}=v''
 $$
 such that for each \(i\), \(0 \leq i \leq n-1\) the vertices
 \(v_{i}\), \(v_{i+1}\) are connected by some line.

 The graph is called connected if and only if any two vertices are connected
 by some path.

A Feynman graph \(\Phi\) is called one particle irreducible (1PI)
if it is connected and can not be done disconnected by removing a
single line.

 Now let us give some definitions necessary to give a definition of the
 Feynman graphs.

 Let \(\Phi=(V_{\Phi},R_{\Phi},f_{\Phi})\) be a Feynman graph and
  \(v \in V_{\Phi} \) be some its vertex. Let
 $$
 \mathcal{P'}_{v}=\bigotimes \limits _{r \in R^{\rightarrow v}} \mathcal{P}
\bigotimes \limits _{r \in R^{\leftarrow v}} \mathcal{P}.
 $$
 Here \(\mathcal{P}\) is a space of polynomials on \(\mathbb{R}^4\), \(R^{\rightarrow
 v}\) is a set of all lines  coming into the vertex \(v\) and \(R^{\leftarrow v}\)
 is a set of all lines coming from the vertex \(v\). We can see at \(\mathcal{P'}_{v}\)
 as at polynomials \(\mathbf{p}(p_1,...,p_n)\)
 of \(n\) four-vectors arguments. \(n= \sharp R^{\rightarrow v}+\sharp R^{\leftarrow
 v}\).
 Consider the subspace \(
  \mathcal{N}_{v}\) of the space \(\mathcal{P'}_{v}\) of all
  polynomials \(\mathbf{p}(p_1,...,p_n)\) such that \(\mathbf{p}(p_1,...,p_n)=0\) if \(
  p_1+...+p_n=0\). The space of vertex operators by definition is
  $$
  \mathcal{P}_{v}=\mathcal{P'}_{v}/ \mathcal{N}_{v}.
  $$

  \textbf {Definition.} The Feynman diagram is a pair
  $$
  \Gamma=(\Phi_\Gamma ,\varphi_\Gamma ),
  $$
  where \(\Phi_\Gamma \) is a Feynman graph and
  \(\varphi_\Gamma\) is a map which assigns to each vertex \(v\)
  an element \(\varphi(v)\) of \(\mathcal{P}_{v}\).

 \textbf {Definition.} The Feynman diagram \(\Gamma=(\Phi_\Gamma ,\varphi_\Gamma )\)
  is called connected, one particle irreducible, if the
  corresponding Feynman graph \(\Phi_\Gamma\) connected, one particle irreducible
  respectively.

 \textbf {Definition.} Let \(\Gamma=(\Phi ,\varphi )\) be an one
 particle irreducible diagram. \(\Phi=(V,R,F)\). Power of
 divergence of \(\Gamma\) by definition is a number
 $$
 \Omega_\Gamma=\sum \limits_{v \in V} (\rm deg \mit \varphi_\Gamma (v)
 -4)+\sum \limits_{r \in R_{in} } (4-2) +4.
 $$

\textbf{The Feynman Propagator.} By definition Feynman propagator (Euclidean) is a map
 $$
 \Delta:\mathbf{R^{4}}\rightarrow\mathbf{R},
 $$
 $$
 \Delta: p \mapsto \Delta(p)=\frac{1}{p^{2}+m^{2}},
 $$
 $$
 m>0.
 $$
 Note, that
 $$
 \Delta(p)= \int \limits_{0}^{+\infty} e^{-\alpha(p^{2}+m^{2})}
 d\alpha.
 $$
 This representation is called the Schwinger de Witt representation.\newline

 Define an analytically regularized propagator as follows
 $$
\Delta^{z}(p)= \int \limits_{0}^{+\infty} \alpha^{z}
e^{-\alpha(p^{2}+m^{2})} d\alpha.
$$
It is easy to see that
$$
\Delta^{z}(p)=\frac{\Gamma(1+z)}{(p^{2}+m^{2})^{1+z}}.
$$

\textbf{The Feynman Amplitudes.} Let \(\Phi_\Gamma\) be a Feynman
diagram. Let \(n=\sharp R_{ext}\) be a number of external lines.
Let \(p_1,...,p_n\) be momenta of particles coming into (or from)
the diagram. The Feynman amplitude is a function of
\(p_1,...,p_n\) defined as follows,
\begin{eqnarray}
U^{z}_{\Gamma}(p_1,...,p_n)\delta(p_1+,...,p_n)\nonumber\\
=\int \prod \limits_{r \in R_{in}} d p'_r \prod \limits_{r \in
R_{in}} \Delta^{z}(p'_r) \prod \limits_{v \in V} \varphi_{\Gamma}(
\pm p_{r_v^1}...\pm p_{r_v^{k_{v}}}) \delta( \sum
\limits_{r\rightarrow v} p'_r - \sum \limits_{r\leftarrow v}
p'_r). \label{1}
\end{eqnarray}
Let us describe the basis elements of this formula. \newline
  a) The integration is over all internal momenta. \newline
  b) The symbol \(r\rightarrow v\) means that the line \(r\) comes into the vertex \(v\),
     the symbol \(r\leftarrow v\) means that the line \(r\)
     comes from the vertex \(v\). \newline
  c) Let \(r \in R_{ext}\). By definition \(p'_r=p_{k_r}\) ( \(k_r\) is a number of
  the line \(r\)) if \(r\) comes into \(v\), and
     \(p'_r=-p_{k_r}\) if \(r\) comes from \(v\).\newline
  d) \(k_v\) is a number of lines coming into the vertex \(v\) or coming from the vertex v.
  The lines \(r_v^i\) are the lines coming into the vertex \(v\) or coming from the vertex v.
  We take the sign \(+\) in arguments of \(\varphi_{\Gamma}\) if the
  corresponding line comes into the vertex and the sign - if the
  corresponding line comes from the vertex.

  We will prove below that the integral (\ref{1}) converges and defines an
  enough many times differentiable function of \(p_1,...,p_n\)  if \(\rm Re \mit\; z\)
  is enough large. Let
  $$
  \newline \Omega_{\Gamma}^z=\sum \limits_{v \in V} (\rm deg \mit \varphi_\Gamma (v)
  -4)+\sum \limits_{r \in R_{in} } (4-2(1+z)) +4.
  $$
  \section {The Bogoliubov --- Parasiuk R-operation}
  Let  \(\Gamma=(\Phi_\Gamma ,\varphi_\Gamma )\),
 \(\Phi=(V,R,f)\).

 Let \(V'\) be a subset of \(V\). Let \( \tilde{R'}\) be a subset of \(R\)
 satisfying the following condition. For each line \(r\) of \(\tilde{R'}\) there exist
 the vertices \(v,\;v' \in V'\) such that \((v,v')=f(r)\).

 Let \(R'=\tilde{R'}\cup\tilde{R''}\), where
 \(\tilde{R''}\subset (R \setminus \tilde{R'})\times \{+,-\} \)
 which consists of all pairs \((r,+)\) such that \(r\) comes into \(V'\) and all pairs \((r,-)\)
 such that \(r\) comes from \(V'\) (\(r \in R \setminus \tilde{R'}\)).

 Consider the graph \(\Phi_\gamma=(V',R',f')\), where \(V'\) and  \(R'\)
  are just defined, and

   \(f'(r)=f(r)\) if \(r \in \tilde{R'}\)

 \(f'(r)=(v,+)\) if \(r \in \tilde{R''}\), \(r=(r',+)\) for some line \(r' \in R\) and
 \(r'\) comes into \(v\).

\(f'(r)=(v,-)\) if \(r \in \tilde{R''}\), \(r=(r',-)\) for some
line \(r' \in R\) and
 \(r'\) comes from \(v\).

 Let \(\gamma =(\Phi_\gamma,\varphi_\gamma)\) be a diagram, where
 \(\varphi_\gamma\) is a restriction of \(\varphi_\Gamma\) on
 \(V'\). If \(\gamma\) is one particle irreducible diagram than \(\gamma\) is called an
 one particle irreducible subdiagram of \(\Gamma\).

 Let \(\Gamma\) be one particle irreducible diagram and \(\gamma\) be its one particle
 irreducible subdiagram. Let \(C_\gamma(p_1,...,p_l)\) be a
 polynomial of external momenta \(p_1,...,p_l\) of \(\gamma\).

 Define \(C_\gamma \star U_\Gamma (p_1,...,p_n)\) by the following way:
 consider a diagram \(\Gamma/\gamma:=(\Phi_{\Gamma/\gamma},
 \varphi_{\Gamma/\gamma})\), where \(\Phi_{\Gamma/\gamma}\) is so called quotient graph
 obtained by replacing of \(\Phi_\gamma\) by a vertex \(v_0\). \(
 \varphi_{\Gamma/\gamma}=\varphi_\Gamma(v)\) if \(v\neq v_0\) and
 \(\varphi_{\Gamma/\gamma}(v_0)(p_1,...,p_l)= C_\gamma(p_1,...,p_l)\). Put by definition
  \(C_\gamma \star U_\Gamma (p_1,...,p_n)=U_{\Gamma/\gamma} (p_1,...p_l)\).

  We can define \(C_{\gamma_1}\star...\star C_{\gamma_n}\star U_\Gamma\) by a similar
  way. Here \(\gamma_1\),...,\(\gamma_n\) are one particle irreducible
 subdiagrams of \(\Gamma\) such \(\forall (i,j)\) \( i\neq j\)
 \(i,j=1,...,n\)
 the diagrams \(\gamma_i\), \(\gamma_j\) do not intersect.\newline

 Now let us define the R-operation. Let us suppose that our theory satisfy
 the following condition: at each order of perturbation theory
 there exists only a finite number of diagrams. Let us define the amplitudes
 \(R_\Gamma^z (p_1,...,p_n)\), \(C_\Gamma^z (p_1,...,p_n)\) by the following recurrent relation.
 Suppose that for each diagram \(G\) such that the number of its
 vertices less than \(n\) the amplitudes \(R_\Gamma^z\) and \(C_\Gamma^z\) are defined.
 To define the amplitudes \(R_\Gamma^z\) and \(C_\Gamma^z\) at \(n\)-th order,
 we must move all diagrams of order \(n\) starting from the diagrams
 with maximal numbers of external lines and going to the diagrams with
 minimal number of external lines and use the following formulas.
\begin{eqnarray}
C^z_\Gamma=-\mathbf{T}(U^{z}_\Gamma + \sum \limits_{\gamma_1
\star...\star \gamma_n \subset \Gamma} C^{z}_{\gamma_1}
\star...\star C^{z}_{\gamma_n} \star U^{z}_\Gamma ), \nonumber \\
R^z_\Gamma=(1-\mathbf{T})(U^{z}_\Gamma + \sum \limits_{\gamma_1
\star...\star \gamma_n \subset \Gamma} C^{z}_{\gamma_1}
\star...\star C^{z}_{\gamma_n} \star U^{z}_\Gamma ). \label{RR} \nonumber \\
\end{eqnarray}
Here \(\mathbf{T}\) is a n operator which to each Laurant series
with the center at zero
\(\frac{a_{-n}}{z^n}+...+\frac{a_{-1}}{z}+a_0+a_1z+...\) assigns
its pole part \(\frac{a_{-n}}{z^n}+...+\frac{a_{-1}}{z}+a_0\).
\(\subset\) means the strong inclusion.\newline

 \textbf{Theorem (Bogoliubov --- Parasiuk).}
 For each diagram \(\Gamma\) there exists a polynomial \(C_\Gamma^z (p_1,...,p_n)\)
 of external momenta such that \(\rm deg \mit C_\Gamma^z
 (p_1,...,p_n)\leq \Omega_\Gamma\), its coefficient are the
 polynomials on the inverse powers of \(z\), and the following
 conditions
 are satisfied.
 \newline
 a)
\begin{eqnarray}
(U^{z}_\Gamma + \sum \limits_{\gamma_1 \star...\star \gamma_n
\subset \Gamma} C^{z}_{\gamma_1} \star...\star C^{z}_{\gamma_n}
\star U^{z}_\Gamma ) \nonumber
\end{eqnarray}
has an unique analytical continuation into the point \(z=0\).
\newline
b) The recurrent relations (\ref{RR}) holds.\newline

 Our aim is to prove this theorem. Let us give some useful definitions.\newline
 Let \(n \in \mathbf{Z}\). Put by definition \( \mathbf{R_{+}}:= \{x \in \mathbf{R}\mid x
\geq 0 \}\), and let \(\alpha_1,....\alpha_n\) are the coordinates
on
 \(\mathbf{R_{+}}^{n}\), \(\alpha_i \geq 0\).

 The space of all test functions on \(\mathbf{R}_{+}^{n}\) \({S}(\mathbf{R}_{+}^{n})\)
 consists by definition of all smooth functions on
 \(\mathbf{R_{+}}^{n}\) which decay faster than any inverse polynomial
 on \(\alpha_1+...+\alpha_n\) with all its derivatives as
 \(\alpha_1+...+\alpha_n\) tends to infinity.

 This space is a Frechet space with respect to the following set
 of seminorms:
 \begin{equation}
 {\|f\|}_n = \sup_{ \{ \vec{\alpha} \in \mathbf{R_{+}^{n}},| \vec{m}| \leq n \}}
 |(1+|\vec{\alpha}|)^{n} f^{( \vec{m})}|,
\end{equation}
where \(\vec{\alpha}=(\alpha_1,...,\alpha_n)\),\newline
  \(\vec{m}=(m_1,...,m_n)\) \(m_i \in \mathbf{Z}\),\newline
  \(|\vec{m}|=m_1+...+m_n\), \newline
  \(|\vec{\alpha}|=\alpha_1+...+\alpha_n\),\newline
  \(f^{(\vec{m})}=\frac{{\partial}^{|\vec{m}|} f}{{\partial}^{m_1}....{\partial}^{m_n}}\).\newline
Denote by \( S'({\mathbf{R}}^{n}_{+})\) the topological dual of \(
S({\mathbf{R}}^{n}_{+})\). The space \( S'({\mathbf{R}}^{n}_{+})\)
is called the space of distributions on \({\mathbf{R}}^{n}_{+}\).
If \(f \in S'({\mathbf{R}}^{n}_{+})\) then there exist such \(m
\in \mathbf{Z} \), \(C \in \mathbf{R}_+\) as
\begin{equation}
|\langle f,g\rangle| \leq C \|g \|_m
\end{equation}
\( \forall g \in S'({\mathbf{R}}^{n}_{+})\) \newline (The Laurent
Swartz theorem).

 Let us denote by \({S'_m}({\mathbf{R}}^{n}_{+})\) the space of all distributions which are
continuous with respect to the norm \(\|\|_m\). Let us introduce
in \( {S'_m}({\mathbf{R}}^{n}_{+})\) the norm \(\|\|'_m\) as
follows:
\begin{eqnarray}
\|f\|'_m=\inf\{C||\langle f,g\rangle| \leq C \|g \|_m\;\forall g
\in S(\mathbf{R}_+^n)\}.
\end{eqnarray}
We have \newline
 \(S'_0 \subset S'_1 \subset...\subset S'_n. \) \newline
 All injection are continuous and \( \bigcup \limits_{i=0}^{\infty}
  S'_i =S'\).\newline
 Let \( S_m\) be a completion of \(S\) with respect to the norm \(\|\|_m\).
 We can see at \( S'_m\) as at dual of the
  \(S_m\). We have \(S_0 \supset S_1 \supset...\supset S_n \). All injection are continuous.
 Let us fix an one particle irreducible diagram \(\Gamma\).
 Let us define the Feynman amplitude as a distribution on
  \(({\mathbf{R}}_+^n) :=
  \mathbf{R}_{+}^\Gamma\) (n is a number of elements of \(R_{in}\)  by the formula
\begin{eqnarray}
U^{z}_{\Gamma}(p_1,...,p_n)(\vec{\alpha})\delta(p_1+,...,p_n) \nonumber\\
=\int \prod \limits_{r \in R_{in}} {\alpha_r}^{z} \prod \limits_{r
\in R_{in}} d p'_r  \prod \limits_{v \in V} \varphi_{\Gamma}( \pm
p_{r_v^1}...\pm p_{r_v^{k_{v}}}) \delta( \sum
\limits_{r\rightarrow v} p'_r - \sum \limits_{r\leftarrow v}
p'_r)\prod \limits_{r \in R_{in}} e^{-\alpha_r {p'}^2_r}.
\end{eqnarray}
\section{Estimates of the Feynman amplitudes }
Now we will prove that if \(\rm Re \mit\; z\) is enough large the
Feynman amplitude is an integrable function. Consider the Feynman
graph \(\Phi\) corresponding to \(\Gamma\). Let \(\Phi'_{\Delta}\)
be its some maximal tree, and \(R'_{in}\) is a set of all internal
lines of \(\Phi\) which do not belong to \(\Phi'_{\Delta}\). The
set of momenta \(p'_r\) \(r \in R'_{in}\) (after remooving of all
\(\delta\)-functions) determines uniquely all others momenta. The
momenta \(p'_r\) \(r \in R'_{in}\) are called the loop momenta. We
have
\begin{eqnarray}
U^{z}_{\Gamma}(p_1,...,p_n)(\vec{\alpha})\delta(p_1+,...,p_n)= \nonumber\\
\int \prod \limits_{r \in R_{in}} {\alpha_r}^{z} \prod \limits_{r
\in R'_{in}} d p'_r  \prod \limits_{v \in V} \varphi_{\Gamma}( \pm
p_{r_v^1}...\pm p_{r_v^{k_{v}}}) \prod \limits_{r \in R_{in}} e^{-
\alpha_r {p'}^2_r}.
\end{eqnarray}
The integrand has the form of polynomial multiplied by the Gauss
function.\newline

 \textbf{Lemma.} Let \(P\) be a polynomial of \(x_1,...,x_n\; x_i \in \mathbb{R},i=1,...,n\)
 and \(Q\) be a positive definite quadratic form of  \(x_1,...,x_n\). Then

\begin{equation}
| \int |P| e^{-Q} d x_1...d x_n| \leq
C(\frac{1}{{\lambda_{min}}^{\frac{n+deg P}{2}}}+1),
\end{equation}
where \(\lambda_{min}\) is a minimal eigenvalue of \(Q\), and the
constant \(C\) do not depend of \(P\).

 \textbf{Proof.}
\newline
To prove the lemma it is enough to consider the integral
\begin{equation}
 \int |x|^{p} e^{-\lambda x^2} d x
\end{equation}
But for this case the lemma is evidence.

 So we must to find the lower estimate for the quadratic form \( \sum
\limits_{r} \alpha_r
 {p'_r}^{2} \). We suppose that all momenta \({p'}\) are defined by the loop
 momenta, and all external momenta are supposed to be equal to zero.

  \textbf{Lemma.} There exists an absolute constant \(C\) (\(C\)
  do not depend of \(\alpha_r\)) such that:
\begin{equation}
Q \geq C({\sum {p'_r}^2}){ \rm min \mit \{\alpha_r\}}.
\end{equation}

 \textbf{Proof.}

Let us use the following electrotechnical analogy. The Feynman
graph corresponds to the electrical scheme. Momenta \(p_r\)
corresponds to the currents. The parameters \(\alpha_r\)
corresponds to the resistances. The low of momenta conservation
corresponds to the first Kirhhoff low. According to the Joule ---
Lenz low the heat generating by resister number \(r\) is equal to
\(Q_r=\alpha_r {p'}^{2}_{r}\). Let \(Q\) be a total heat
generating by the scheme. Suppose that \(Q \leq 1\). The heat
generating by resister number \(r\) less or equal than \(Q\).
Therefor \( {p'}^2_r \leq \frac{1}{\alpha_r} \). The lemma is
proved.
\newline

 \textbf{Lemma.} Let \(p_1,...,p_n\) be external momenta. Let
 \({q}_r\) be internal momenta corresponding to the minimum of the quadratic form \(Q\).
 Variables \({q}_r\) satisfy to the following inequality (maximum
 principle):
 \begin{equation}
 \|q_r\|\leq C\|{p}_r\|
 \end{equation}
 for some constant \(C\) that does not depend of \(\alpha_r \) and \(p'_1,...,p'_n\). It is possible to use
 an arbitrary norms in this inequality.\newline

\textbf{Proof.}

Let \(N\) be a maximal number of lines coming into (or from) a
vertex, and

 \(P:=\max \limits_{r \in R_{ext}} {|p_r|}\), \( R=\sharp R_{in}\). Put \(C=2PN^{R}\).
 Let us prove that for all \(r \in R_{in}\) \(q_r \leq
 C\).\newline
 Note that \(q_r\) satisfy the condition: the fall of voltage at each closed contour is
 equal to zero (The second Kirhoff low). But according to the Ohm low the fall of
 voltage at the resister \(r\) is equal to \(\alpha_r q_r\). Suppose
 that there exists a line
 \(r_1 \in R_{in}\) such that \(|q_{r_1}|>C\). Let \(v\) be a
 vertex such that a line \(r_1\) coming into \(v\). Let us consider other lines coming into the vertex \(v\).
 It is evidence that there exists an internal line \(r_2\) \(r_1\neq r_2\)
 such that \(|q_{r_2}|>2PN^{R-1} \). This
 fact follows from the first Kirhoff low. By the same way for the
 line \( r_2\) we find the line \(r_3\) such that:
 \(|q_{r_3}|>2PN^{R-2} \) e.t.c. After at most \(R\) steps we find lines \(r_1,...,r_m\) such that
 for each \(i=1,...,m-1\) the lines \(r_i\) and \(r_{i+1}\) ends at a common vertex,
 the current flows in the same direction at each line and
 there exists a vertex \(v\) and number \(i=1,2,...m\) such that
 the lines \(r_i\) and \(r_m\) ends at the same vertex. So we have
 a closed contour such that the fall of voltage at this contour is
 not equal to zero. This contradiction with the second Kirhhof low
 proves the lemma.

We find from two previous lemmas that for fixed \(p_1,...,p_n\)
there exists a constant \(C\) such that
 \begin{eqnarray}
 |U_{\Gamma} (\vec{\alpha})(p_1,...,p_n)| \nonumber\\
 \leq C (\prod \limits_{r \in R_in} \alpha_r^z)\{\frac{1}{(\rm min \mit \{
 \alpha_r\})^{2(\sharp R_{in}-\sharp V +1)+1/2(\sum \limits_{v \in
V} deg
 \varphi_v)}}+1\} \nonumber \\
=C (\prod \limits_{r \in R_{in}} \alpha_r^z)\{\frac{1}{(\rm min
\mit
 \{\alpha_r\})^{1/2\Omega_\Gamma+\sharp R_{in}}}+1\}.
 \end{eqnarray}
 We see that if \( \rm Re \mit z >1/2\Omega_\Gamma+\sharp R\) then
\( U_{\Gamma} (\alpha)(p_1,...,p_n)\) is a continuous function of
variables \(p_1,...,p_n\). It is also clear that if
\(\rm Re \mit\; z
>\frac{1}{2} \Omega_\Gamma+\sharp R+m\) than \( U_{\Gamma}
(\alpha)(p_1,...,p_n)\) is \(m\)-times continuously differentiable
function.

 \textbf{Homogeneity of \(U_{\Gamma} (\vec{\alpha})(p_1,...,p_n).\)}
Let us introduce an operation \(\Lambda_\lambda\) which acts by
the formula
 \begin{equation}
(\Lambda_\lambda U_{\Gamma}) (\vec{\alpha})(p_1,...,p_n):=
U_{\Gamma} (\lambda
\vec{\alpha})(\frac{p_1}{{\lambda}^{1/2}},...,\frac{p_n}{{\lambda}^{1/2}}).
\end{equation}
It is clear that if all \(\varphi_v\) are homogeneous then
\begin{equation}
\Lambda_\lambda
(U_{\Gamma})=\frac{1}{{\lambda}^{1/2\Omega_\Gamma+\sharp
R_{in}(1-z)}}U_{\Gamma}=\frac{1}{{\lambda}^{1/2\Omega_\Gamma^{z}+\sharp
R_{in}}}U_{\Gamma}.
\end{equation}
\section{Local amplitudes, theorems II, III}

\textbf{Definition.} Let \(\Gamma\) be a Feynman diagram.
Distribution \(C_{\Gamma} (\vec{\alpha})(p_1,...,p_n)\) is called
local if it is a linear combination of \(\delta\)-function and its
derivatives
(\(\delta(\vec{\alpha})=\delta(\alpha_1)...\delta(\alpha_n)\))
with polynomial on external momenta coefficients, homogenous with
respect \(\Lambda_\lambda\) of degree \(1/2\Omega_\Gamma+\sharp
R_{in}\). In other words the function is local if it has the form
\begin{equation}
C_{\Gamma} (\vec{\alpha})(p_1,...,p_n) =\sum \delta^{(\vec{m})}
(\vec{\alpha}) P_{\vec{m}} (p_1,...,p_n), \rm and \mit
\end{equation}
\begin{equation}
\sharp R_{in}+|\vec{m}|+1/2 \rm deg \mit
P_{\vec{m}}=(1/2)\Omega_\Gamma+\sharp R_{in}.
\end{equation}
Let \(U_\Gamma\) be a Feynman amplitude, \(\gamma\) be a one
particle irreducible diagram, and \(C_\gamma\) be a local
amplitude. Let us define \(C_\gamma \star U_\Gamma\).

Let \((\Gamma/\gamma)^{\vec{m}}\) be a Feynman diagram such that
 \(\Phi_\Gamma /\Phi_\gamma\) be a corresponding Feynman graph,
 and the function \(\varphi_{(\Gamma/\gamma)^{\vec{m}}}\) is defined as follows:
 if \(v\) is not a vertex, obtained by removing of \(\Phi_\Gamma\)
 by a point, then \(\varphi_{(\Gamma/\gamma)^{\vec{m}}}=\varphi_\Gamma\),
 otherwise \(\varphi_{(\Gamma/\gamma)^{\vec{m}}}=P_{\vec{m}}(p_1,...,p_{n}).\)
 Let
\begin{equation}
(C_\gamma \star U_\Gamma)(\vec{\alpha})(p_1,...,p_n)= \sum
\limits_{\vec{m}}
 U_{(\Gamma/\gamma)^{\vec{m}}}(p_1,...,p_n) \bigotimes
 \limits_{\Gamma/\gamma} \delta^{(\vec{m})} (\vec{\alpha}).
 \end{equation}
 Here the tensor product \(\bigotimes
 \limits_{\Gamma/\gamma}\) has the following meaning. If
 \newline
 \(f(\vec{\alpha})=\bigotimes
 \limits_{r \in R^{in}_{\Gamma/\gamma}} f_r (\alpha_r)\) is a
 distribution on \(R^{\Gamma/\gamma}_+\) and \newline
  \(g(\vec{\beta})=\bigotimes
 \limits_{r \in R^{in}_{\gamma}} g_r (\beta_r)\) is a
 distribution on \(R^{\gamma}_+\) \newline
 then \(f\bigotimes_{\Gamma/\gamma} g(\vec{\gamma})\) is a distribution on
 \(R^{\Gamma}_+\) equal to \(\bigotimes \limits_{r \in R_{\Gamma}}
 l(\gamma_r)\), where \(l(\gamma_r)=f(\gamma_r)\), if \(r \in
 R^{in}_{\Gamma/\gamma}\) and  \(l(\gamma_r)=g(\gamma_r)\), if \(r \in
 R^{in}_{\gamma}\). \newline

 Let  \(\gamma_1\),...,\(\gamma_m\) \(i=1,...,m\)  be a set of one particle
 irreducible diagrams such that \(\forall i,j=1,...,n,\; i\neq j\)
 the set of vertices of diagrams \(\gamma_i\), \(\gamma_j\) are
 not intersecting. Let \(C_{\gamma_i}\) (\(i=1,...,m\)) be some local
 amplitudes. As above we can define the distribution
 \(C_{\gamma_1} \star...C_{\gamma_n} \star U_\Gamma\).
 Now let us formulate our main theorem. The Bogoliubov --- Parasiuk theorem
 follows from this theorem immediately. \newline

\textbf{Theorem II.} For each one particle irreducible diagram
\(\Gamma\) there exists a local distribution
\(C_\Gamma(\vec{\alpha})(p) \in S'({\mathbf{R}}_+^{\Gamma})\),
that is at the same time is a polynomial on \(\frac{1}{z}\), such
that:
\newline
 a)\begin{eqnarray}
(U^{z}_\Gamma + \sum \limits_{\gamma_1 \star...\star \gamma_n
\subset \Gamma} C^{z}_{\gamma_1} \star...\star C^{z}_{\gamma_n}
\star U^{z}_\Gamma ) \nonumber
\end{eqnarray}
has an unique analytical extension in some punctured neighborhood
of the point \(z=0\).
\newline
b)
\begin{eqnarray} C^z_\Gamma=-\mathbf{T}(U^{z}_\Gamma + \sum
\limits_{\gamma_1 \star...\star \gamma_n \subset \Gamma}
C^{z}_{\gamma_1}
\star...\star C^{z}_{\gamma_n} \star U^{z}_\Gamma ) \nonumber \\
\end{eqnarray}
 \(\mathbf{T}\) is an operator, which to each function holomorphic in some
punctured neighborhood of zero assigns its pole part.
\newline
Before we begin to proof this theorem let us give some preliminary
definitions.\newline Let \({\Upsilon'}_{\Gamma}^{m,n}\) be a space
of all functions \(f\) belongs to \({S'}^{m}_{\Gamma}\) such that
their are smooth functions of \(p_1,...,p_f\) (i.e. \(\forall
 \varphi \in S^{m}_{\Gamma}\) \(\langle f,\varphi\rangle\) ---
 is a smooth function of external momenta) and,
 \begin{equation}
 \|(\Lambda_\lambda f)(\vec{\alpha})(p)\|'_{m}\leq
 C^{f}_{\varepsilon}{\lambda}^{-\varepsilon-n}
 \end{equation}
 if \(|p_1|^2+...+|p_f|^2\leq1\), \(\lambda \leq 1\)
 \(\forall \varepsilon >0\). \(C_{\varepsilon}^{f}\) depend of
 \(f\) and \(\varepsilon\).\newline
 Now let us investigate the behavior of
 \(\langle f(\vec{\alpha})(p),g(\vec{\alpha})\rangle\)
 at large momenta. If \( \|p\|^2\leq 1\), then
 \begin{equation}
 \langle f(\vec{\alpha})(p),g(\vec{\alpha})\rangle \leq
 C^{f}_{\varepsilon}\|g\|_{m}.
\end{equation}
Suppose that \(\|p\|>1\). Let
\(\frac{1}{\lambda^{\frac{1}{2}}}=\|p\|\) and let us introduce
\(p'\) such, that \(p=\frac{p'}{\lambda^{\frac{1}{2}}}\).
\begin{eqnarray}
|\langle f( \vec{\alpha})(p),g(\vec{\alpha})\rangle|\nonumber\\
=\langle
(\Lambda_{\lambda}f)(\frac{\vec{\alpha}}{\lambda})(\lambda^{\frac{1}{2}}p),g(\vec{\alpha})\rangle
\leq \langle (\Lambda_\lambda
f)(\vec{\beta})(p'),g(\lambda\vec{\beta}))\rangle{\lambda}^{\sharp R_\Gamma} \nonumber\\
\leq C_{\varepsilon}^{f} {\lambda}^{\sharp R_\Gamma}
{\lambda}^{-\varepsilon-n}\|g(\lambda\vec{\beta})\|_{{m}}. \nonumber\\
\|g(\lambda \vec{\beta})\|_{{m}} \leq \lambda^{-{m}}
\|g(\vec{\beta})\|_{{m}}.
\nonumber\\
\end{eqnarray}
So we have the following\newline

\textbf{Lemma.} If \(f \in {\Upsilon'}^{m,n}\), \(g \in S^{m}\)
then
 \( |\langle f(\vec{\alpha})(p),g(\vec{\alpha})\rangle| \leq
 C_{\varepsilon}^{f}(1+|p|^{2(\varepsilon+m+n-\sharp
 R_\Gamma)})\|g(\vec{\alpha})\|_m\).

 Let us make a remark. If \(f \in {\Upsilon'}^{m,n}\) and
\(\|(\Lambda_\lambda f)(\vec{\alpha})(p)\|'_{{m}} \leq
 C^{f}_{\varepsilon}{\lambda}^{-\varepsilon-n}\), where \(|p|\leq
 1\) and \(\lambda\leq 1\) then \( \Lambda_\mu (f) \in {\Upsilon'}^{m,n}\)
(\(\mu <1\)) and \( \|(  \Lambda_{\lambda}(\Lambda_\mu
f))(\vec{\alpha})(p) \|'_m \leq
 C^{\Lambda_\mu f}_{\varepsilon} {\lambda}^{-\varepsilon -n}\),
 where
\(C^{\Lambda_\mu
f}_{\varepsilon}=C^{f}_{\varepsilon}{\mu}^{-\varepsilon-n}
\).\newline

We say, that the function \(f_{z} \in {\Upsilon'}^{m,n}\) is a
holomorphic function of \(z\) in the region \(\mathcal{O}\), if \(\forall
p\) \(\forall g\) \(\langle
f_{z}(\vec{\alpha})(p),g(\vec{\alpha})\rangle\) is a holomorphic
function of z in this region.\newline

Now we formulate some theorem. Theorem II follows from this
theorem.

  \textbf{Theorem III}
  For each diagram \(\Gamma\) we can construct an local amplitudes \(C^{z}_{\Gamma}(\vec{\alpha})(p)\),
  such that the following conditions are satisfied:

  a) \(C^{z}_{\Gamma}(\vec{\alpha})(p)\)
   is a polynomial on \(1/z\).

  b) \begin{eqnarray}
(U^{z}_\Gamma + \sum \limits_{\gamma_1 \star...\star \gamma_n
\subset \Gamma} C^{z}_{\gamma_1} \star...\star C^{z}_{\gamma_n}
\star U^{z}_\Gamma ) \nonumber
\end{eqnarray}
has an unique analytical continuation to the punctured neighborhood of
the point \(z=0\).

  c) Amplitudes \(C^{z}_{\Gamma}\) satisfy to the following recurrent
relation:
\begin{eqnarray} C^z_\Gamma=-\mathbf{T}(U^{z}_\Gamma + \sum
\limits_{\gamma_1 \star...\star \gamma_n \subset \Gamma}
C^{z}_{\gamma_1}
\star...\star C^{z}_{\gamma_n} \star U^{z}_\Gamma ).
\end{eqnarray}

Denote by \({}^{(l)}U_\Gamma\) some l-th derivatives of the
Feynman amplitude \(U_\Gamma\). Let

\begin{eqnarray} R^z_\Gamma=(1-\mathbf{T})(U^{z}_\Gamma + \sum
\limits_{\gamma_1 \star...\star \gamma_n \subset \Gamma}
C^{z}_{\gamma_1}
\star...\star C^{z}_{\gamma_n} \star U^{z}_\Gamma ).
\end{eqnarray}

Then

d)

i)The function \(R^z_\Gamma\) is holomorphic for enough large
\(\rm Re \mit\; z\) and holomorphic in the whole open complex
plane except \(D \setminus \{0\}\), where \(D\)
--- some discrete set.

ii)\(R^z_\Gamma\) is a smooth function of external momenta and
 \begin{equation}
{}^{(l)} R^z_\Gamma \in \Upsilon_{\Gamma}^{m,((1/2)
\Omega_{\Gamma} + \sharp R_{\Gamma}-\frac{|l|}{2}+x^{l}_{\Gamma}
{(\rm Re \mit\; z)}^{-})}.
\end{equation}

if \(\rm Re \mit\; z >-\epsilon_\Gamma\). Here \(m\) depends of
\(\Gamma\) \(l\), \(x^{l}_{\Gamma}\) is some positive number,
\(\epsilon_\Gamma\) is a positive number, \((\rm Re \mit\;
z)^{-}:= \rm max \mit(-\rm Re \mit\; z, 0)\). Constants
\(C_{\varepsilon}\) depends continuously on \(z\) in the whole
complex plane except \(D \setminus \{0\}\).

 iii)\(R^z_\Gamma\) can be represented as a sum \newline
 \(R^z_\Gamma=\sum \limits_{\delta}
 {(R^z_\Gamma)}^{(\delta)}\),\newline
 where \({(R^z_\Gamma)}^{(\delta)}\) are homogenous
 amplitudes with respect \(\Lambda_{\lambda}\); \newline
 \(\Lambda_{\lambda} {(R^z_\Gamma)}^{(\delta)}=
 {(R^z_\Gamma)}^{(\delta)}{\lambda}^{(-\frac{\Omega_\Gamma}{2}+
 \sharp R + y^{\delta}_{\Gamma}
 z)},
\) (\(y^{\delta}_{\Gamma}>0\)) and meromorphic on  \(z\). \({(R^z_\Gamma)}^{(\delta)}\) has poles
only at points of the set \(D\).

iv) \(U_\Gamma^z\) is a meromorphic function with poles belongs
 to \(D\).
\section{Beginning of the proof of theorem iii. Radial integration}
 We will prove the previous theorem. Our nearest goal is for all
 \(\lambda > 0 \) to define the distribution \( C_\gamma \star
 U_\Gamma (\vec{\alpha})(p) \delta(|\vec{\alpha}|-\lambda)\).

 Here \(\Gamma\) is one particle irreducible diagram, and
 \(\gamma\) is its one particle irreducible subdiagram. \(C_\gamma
 \) is a local amplitude. Let \( \delta_\kappa (x)
 =\frac{1}{\kappa} \chi( \frac{x}{\kappa})\), where \(\chi\) is a smooth
 function such that, \(\chi > 0\), \( \int \chi(x) dx=1\) ,
 \( \rm supp \mit \chi \in [-\frac{1}{2}, \frac{1}{2}]\) \newline

 \textbf{Lemma.} For enough large \(\rm Re \mit\; z\) the following limit
 exists in the sense of distributions:
 \begin{equation}
 \lim \limits_{\kappa=+0} C_\gamma \star
 U_\Gamma (\vec{\alpha})(p) \delta_\kappa (|\vec{\alpha}|-\lambda-\kappa).
 \end{equation}

 \textbf{Proof.}
 It is enough to prove that \(\forall\; \Gamma\) and \(\vec{m}\)
 there exists a limit
 \begin{equation}
 \lim \limits_{\kappa=+0}
 U_\Gamma (\vec{\alpha})(p) \otimes \delta^{(\vec{m})}(\vec{\beta})
 \delta_\kappa (|\vec{\alpha}|+|\vec{\beta}|-\lambda-\kappa).
 \end{equation}
Let us find a value of this distribution at
\(\Phi(\vec{\alpha},\vec{\beta})\). By definition of tensor
product of distributions we find
\begin{eqnarray}
I_\kappa=\langle U_\Gamma (\vec{\alpha})(p) \otimes
\delta^{(\vec{m})}(\vec{\beta}) \delta_\kappa
(|\vec{\alpha}|+|\vec{\beta}|-\lambda-\kappa),\Phi(\vec{\alpha},\vec{\beta})\rangle\nonumber\\
=\langle \delta^{(\vec{m})}(\vec{\beta})\langle  U_\Gamma
(\vec{\alpha})(p) \delta_\kappa
(|\vec{\alpha}|+|\vec{\beta}|-\lambda-\kappa),\Phi(\vec{\alpha},\vec{\beta})\rangle.
\end{eqnarray}
Let \(\eta_\lambda (\vec{\beta})\) be a function which is equal to
1 in some neighborhood of zero and equal to zero if
\(|\vec{\beta}| \geq \frac{\lambda}{2}\). We have
\begin{eqnarray}
I_\kappa= \langle \delta^{(\vec{m})}(\vec{\beta})\eta_\lambda
(\vec{\beta}) \langle U_\Gamma (\vec{\alpha})(p) \delta_\kappa
(|\vec{\alpha}|+|\vec{\beta}|-\lambda-\kappa),\Phi(\vec{\alpha},\vec{\beta})\rangle.
\end{eqnarray}
The function \(U_\Gamma (\vec{\alpha})\) is enough times
differentiable if \(\rm Re \mit\; z\) is enough large, and for each
positive integer \(n\) there exists a real number \(A\) such that
\(U^{\vec{m}}(\vec{\alpha})=0\) if \(\rm Re \mit\; z>A\),
\(|\vec{m}|\leq n\) and for some \(i=1,...,\sharp R\)
\(\alpha_i=0\). The function
\begin{equation}
\eta_\lambda (\vec{\beta}) \langle U_\Gamma (\vec{\alpha})(p)
\delta_\kappa
(|\vec{\alpha}|+|\vec{\beta}|-\lambda-\kappa),\Phi(\vec{\alpha},\vec{\beta})\rangle
\end{equation}
tends to
\begin{equation}
\eta_\lambda (\vec{\beta}) \int d{\vec{\alpha}}
  U_\Gamma (\vec{\alpha})(p)
\delta
(|\vec{\alpha}|+|\vec{\beta}|-\lambda)\Phi(\vec{\alpha},\vec{\beta})=:
f(\beta)
\end{equation}
with respect the norm \(\|\|_m\). So the limit \(\lim
\limits_{\kappa\rightarrow0} I_\kappa\) exists and equal to
\(\langle\delta^{(m)}(\vec{\beta}),f(\vec{\beta})\rangle\). The
lemma is proved.
\newline
 The aim of the following lemma is to extract radial integration
 in the following expression

\begin{equation}
\langle C_\gamma\star
U_\Gamma(\vec{\alpha})(p),\Phi(\vec{\alpha})\rangle.
\end{equation}

 \textbf{Lemma.} If \(\rm Re \mit\; z\) is enough large then
\begin{eqnarray}
\langle C_\gamma\star
U_\Gamma^z,(\vec{\alpha})(p),\Phi(\vec{\alpha})\rangle\nonumber\\
=\int \limits_{0}^{\infty} d\lambda \langle C_\gamma\star
U_\Gamma^z(\vec{\alpha})(p),\delta(\lambda-|\vec{\alpha}|)\Phi(\vec{\alpha})\rangle.
\end{eqnarray}

\textbf{Proof.} It is enough to prove the lemma for the
distribution
\begin{eqnarray}
U_\Gamma^z,(\vec{\alpha})\otimes \delta^{\vec{m}}(\vec{\beta})\nonumber\\
\vec{\alpha}=(\alpha_1,...,\alpha_k).
\end{eqnarray}
Let \(\eta_\delta\) has the same meaning as in the previous lemma.
If \(\rm Re \mit\; z\) is enough large
 \(U_\Gamma(\vec{\alpha})(1-\eta_\delta(|\vec{\alpha}|)) \rightarrow
 U_\Gamma(\vec{\alpha})\) in the sense of topology of \(S'\) as \(\delta\rightarrow 0\).\newline
 The tensor product of distributions is separately continuous. So
\begin{eqnarray}
I=\langle U_\Gamma,(\vec{\alpha})\otimes
\delta^{(\vec{m})}(\vec{\beta}),
\Phi(\vec{\alpha},\vec{\beta})\rangle\nonumber\\
 =\lim \limits_{\delta\rightarrow 0}\langle
U_\Gamma(\vec{\alpha})(1-\eta_\delta(|\vec{\alpha}|))\otimes
\delta^{(\vec{m})}(\vec{\beta}),\Phi(\vec{\alpha},\vec{\beta})\rangle.
\end{eqnarray}
Let \(\delta_\kappa\) has the same meaning as in the previous
lemma. Note that
\begin{eqnarray}
\int d\lambda
\delta_\kappa(\lambda-|\vec{\beta}|-|\vec{\alpha}|-\kappa)=1.
\end{eqnarray}
We have by using the standard argumentation based on an
approximation of the integral by means Riemann sums:
\begin{eqnarray}
I=\lim \limits_{\delta\rightarrow 0}\int
\limits_{0}^{\infty}d\lambda \langle
U_\Gamma(\vec{\alpha})(1-\eta_\delta(|\vec{\alpha}|))\otimes
\delta^{(\vec{m})}(\vec{\beta}),\Phi(\vec{\alpha},\vec{\beta})
\delta_\kappa(\lambda-|\vec{\beta}|-|\vec{\alpha}|-\kappa)\rangle.
\end{eqnarray}
We have by using definition of the tensor product of
distributions:
\begin{eqnarray}
I= \lim_{\delta\rightarrow 0}\int
\limits_{0}^{\infty}d\lambda \langle \delta^{(\vec
{m})}(\beta)\langle
U_\Gamma(\vec{\alpha})(1-\eta_\delta(|\vec{\alpha}|)),
\delta_\kappa(\lambda-|\vec{\beta}|-|\vec{\alpha}|-\kappa)
\Phi(\vec{\alpha},\vec{\beta})\rangle.
\end{eqnarray}
Let us fix some positive \(\lambda\).  The following function of
\(\vec{\beta}\)
\begin{eqnarray}
\langle U_\Gamma(\vec{\alpha})(1-\eta_\delta(|\vec{\alpha}|)),
\delta_\kappa(\lambda-|\vec{\beta}|-|\vec{\alpha}|-\kappa)
\Phi(\vec{\alpha},\vec{\beta})\rangle.
\end{eqnarray}
has the following limit in \(S^{m}\) as \(\kappa\rightarrow 0\):
\begin{eqnarray}
\langle U_\Gamma(\vec{\alpha})(1-\eta_\delta(|\vec{\alpha}|))
\delta(\lambda-|\vec{\beta}|-|\vec{\alpha}|),
\Phi(\vec{\alpha},\vec{\beta})\rangle,
\end{eqnarray}
and integrand in (36) has an integrable majorant. Therefore
\begin{eqnarray}
I= \lim \limits_{\delta\rightarrow 0}\int
\limits_{0}^{\infty}d\lambda \langle
\delta^{(\vec{m})}(\beta)\langle
U_\Gamma(\vec{\alpha})(1-\eta_\delta(|\vec{\alpha}|))
\delta(\lambda-|\vec{\beta}|-|\vec{\alpha}|),
\Phi(\vec{\alpha},\vec{\beta})\rangle. \label{36}
\end{eqnarray}
Note that the integrand in the right hand side of the last
equality has the following limit as \(\delta\rightarrow 0\)
\begin{eqnarray}
 \langle \delta^{(\vec{m})}(\vec{\beta})\langle
U_\Gamma(\vec{\alpha})
\delta(\lambda-|\vec{\beta}|-|\vec{\alpha}|),
\Phi(\vec{\alpha},\vec{\beta})\rangle.
\end{eqnarray}
 If \(\rm Re \mit\; z\) is enough large \(U_\Gamma(\vec{\alpha})\) tends to zero
 enough fast with some number of its derivatives as
 \(|\alpha|\rightarrow0\) and the integrand in \ref{36} has an
 integrable majorant. So we can finish the proof of the lemma by using Lebegues theorem.\newline
 It is obvious that these two lemmas are hold for
\begin{equation}
C_{\gamma_1}\star...\star C_{\gamma_n}\star U_\Gamma.
\end{equation}

We have proved this lemmas only for particular case in the purpose
of simplicity.\newline

Suppose that the theorem is proved for all diagrams such that
their have less than \(N\) vertices.

The basis of induction is evidence because all diagrams with one
vertex are tree diagrams.

Let us define
\begin{eqnarray}
\tilde{R}^z_\Gamma=(U^{z}_\Gamma + \sum \limits_{\gamma_1
\star...\star \gamma_n \subset \Gamma} C^{z}_{\gamma_1}
\star...\star C^{z}_{\gamma_n} \star U^{z}_\Gamma ). \nonumber \\
\end{eqnarray}
Let \(g \in S(\mathbf{R}_{+}^{\Gamma})\) and \(\rm Re \mit\; z\) is
enough large. It follows from two previous lemmas that:
\begin{eqnarray}
\langle
\tilde{R}^z_\Gamma(\vec{\alpha})(p),g(\vec{\alpha})\rangle=
 \int \limits_{0}^{\infty} d\lambda \langle
\tilde{R}^z_\Gamma(\vec{\alpha})(p)\delta(\lambda-|\vec{\alpha}|),g(\vec{\alpha})\rangle.
\end{eqnarray}
Let us use the following substitution in the right hand side of
last equation \(\vec{\alpha}=\lambda\vec{\beta}\). We find
\begin{eqnarray}
\langle
\tilde{R}^z_\Gamma(\vec{\alpha})(p),g(\vec{\alpha})\rangle=
 \int \limits_{0}^{\infty} d\lambda \lambda^{\sharp R_\Gamma -1} \langle
\Lambda_\lambda(\tilde{R}^z_\Gamma)(\vec{\beta})(\sqrt{\lambda}p)
\delta(1-|\vec{\beta}|),g(\lambda\vec{\beta})\rangle.\label{45}
\end{eqnarray}
\section{Decomposition of unite. Factorization of R-operation}
Note that the "integration" in brackets  in the right hand side of
(44) is over some simplex \(T\). Baricentric coordinates on it are
\(\beta_1,...,\beta_{\sharp R_\Gamma}\). Now let us construct some
decomposition of unit on \(T\). Let us decompose \(T\) into
\(2^n\) nonintersecting sets \(\mathcal{O}_A\), where \(A\) is a
subset of \( \{1,2,...,n\}\), \(n=\sharp R\).
\begin{eqnarray}
\mathcal{O}_A=\{\vec{\alpha} \in T|a_i>\delta \; \rm if
\mit\; i \notin A, \; a_i < \delta\; \rm if \mit\; i \in A\}.
\end{eqnarray}
\(\delta\) is very small. Note that the set
\(\mathcal{O}_{\{1,2,...,n\}}\) is empty, because
\(|\vec{\alpha}|\equiv 1\) on \(T\).

 Let us now consider the
following closed sets
\begin{eqnarray}
\tilde{\mathcal{O}}_A=\{\vec{\alpha} \in T|a_i\geq
\delta(1-\gamma)\;\rm if \mit\; i \notin A,\;
a_i\leq\delta(1+\gamma)\; \rm if \mit \; i \in A\},
\end{eqnarray}
\(\gamma\) is very small. We have \(\mathcal{O}_A\subset\subset
\tilde{\mathcal{O}}_A\), \(
\tilde{\mathcal{O}}_{\{1,...,n\}}=\emptyset\).
 Let us consider the following sets \( X_A=\{\vec{\alpha}|\alpha_i=0\;\rm if \mit\; i \in
 A \}\). The sets \(X_A\) are closed.
 We have   \(X_A\cap\tilde{\mathcal{O}}_{A'}=\emptyset\), if
 \(A \nsubseteq A'\) and \(X_A\cap \mathcal{O}_{A'}\neq\emptyset\),
 if \(A\subseteq A'\).

 Let \(\eta_A(\vec{\alpha}) \) be a function which is equal to \(1\) on
 \(\mathcal{O}_A\) and which is equal to zero outside the region
 \(\tilde{\mathcal{O}}_A\). It is obvious that \(\sum \limits_A
 \eta_A>0\). Consider the following functions
 \(\tilde{\eta}_A (\vec{\alpha})=\frac{\eta_A(\vec{\alpha})}
 {\sum \limits_A \eta_A(\vec{\alpha}) }\). It is obvious that the set of this
 functions is a decomposition of unit on \(T\). One can think that
 \(\tilde{\eta}(\vec{\alpha})\) can be extended to the decomposition of unit
 in some neighborhood of \(T\) in \(\mathbf{R}_{+}^{n}\). Note
 that \(\tilde{\eta}_{A'}(\vec{\alpha})\) is equal to zero in some
 small neighborhood of \(X_{A}\) if \(A\nsubseteq A'\).\newline
 We can rewrite the right hand side of (\ref{45}) as follows:
\begin{eqnarray}
\langle
\tilde{R}^z_\Gamma(\vec{\alpha})(p),g(\vec{\alpha})\rangle\nonumber\\
=\sum \limits_{A\subset\{1,2,...,n\}}
 \int \limits_{0}^{\infty} d\lambda \lambda^{\sharp R_\Gamma -1} \langle
\Lambda_\lambda(\tilde{R}^z_\Gamma)(\vec{\beta})(\sqrt{\lambda}p)
\delta(1-|\vec{\beta}|),\tilde{\eta}_A(\vec{\beta})g(\lambda\vec{\beta})\rangle.
\end{eqnarray}
Let us introduce the following notations. For each
\(A\subset\{1,2,...,n\}\) assign the subdiagrams
\(\gamma_1^{A},...,\gamma_l^{A}\) as follows. To the set
\(A\subset\{1,2,...,n\}\) corresponds some set of lines. If we
paint these lines in red color we obtain some subgraph in
\(\Phi_\Gamma\). Consider all one particle irreducible components
of this subgraph. Let us complete these components by external
lines. In result we obtain subdiagrams
\(\gamma_1^{A},...,\gamma_l^{A}\), where \(l\) is a number of one
particle irreducible components. We have
\begin{eqnarray}
\langle \tilde{R}^z_\Gamma(\vec{\alpha})(p),g(\vec{\alpha})\rangle
\nonumber\\
=\sum \limits_{A\subset\{1,2,...,n\}} \int \limits_{0}^{\infty}
d\lambda \lambda^{\sharp R_\Gamma -1} \langle
\Lambda_\lambda(\{1+\sum
\limits_{\gamma_1\star....\star\gamma_n\subset\Gamma }
C_{\gamma_1}^{z}\star...\star C_{\gamma_n}^{z}\}\star U_{\Gamma}^z)
(\vec{\beta})(\sqrt{\lambda}p) \nonumber\\
\times
\delta(1-|\vec{\beta}|),\tilde{\eta}_A(\vec{\beta})g(\lambda\vec{\beta})\rangle.
\end{eqnarray}
Note that there absents all terms in the internal sum,
corresponding to sets
\(\gamma_1\star....\star\gamma_n\subset\Gamma \) such that for
some \(i\) the subdiagram \(\gamma_i\) are not contained in
\(\gamma_k^{A}\) for all \(k=1,..,l\). We can factorize the
contribution of all other terms. We have, in evidence notations:
\begin{eqnarray}
\langle \tilde{R}^z_\Gamma(\vec{\alpha})(p),g(\vec{\alpha})\rangle
=\sum \limits_{A\subset\{1,2,...,n\}}
 \int \limits_{0}^{\infty} d\lambda \lambda^{\sharp R_\Gamma
 -1}\nonumber\\
 \langle
\Lambda_\lambda(\{\prod \limits_{\gamma_i^{A}}\{1+\sum
\limits_{\gamma_1\star....\star\gamma_k\subset \gamma_i^{A}  }
C_{\gamma_1}^{z}\star...\star C_{\gamma_n}^{z}\}\star \}
U_{\Gamma}^z
)(\vec{\beta})(\sqrt{\lambda}p) \nonumber\\
\times
\delta(1-|\vec{\beta}|),\tilde{\eta}_A(\vec{\beta})g(\lambda\vec{\beta})\rangle.
\end{eqnarray}
Before analyse this expression let us introduce some new
definitions . Let \(\Gamma\) be one particle irreducible diagram.
Let us introduce the following amplitude
\(U^{z}_{\Gamma}(p)(\vec{\alpha})[q]\) which depends on loop
momenta \(q\) by the following formula:
\begin{equation}
U^{z}_{\Gamma}(p_1,...,p_n)(\vec{\alpha})[q]= \nonumber\\
 \prod
\limits_{r \in R_{in}} {\alpha_r}^{z} \prod \limits_{v \in V}
\varphi_{\Gamma}( \pm p_{r_v^1}...\pm p_{r_v^{k_{v}}}) \prod
\limits_{r \in R_{in}} e^{-\alpha_r {p}^2_r}.
\end{equation}
It is supposed at last formula that momenta in the right hand side
of this formula are expressed trough \(q\). To point out the fact
that some expression depends on \(q\) we use the symbol \([q]\).
Analogously we can define:
\begin{equation}
(C_\gamma \star U_\Gamma )(\vec{\alpha})(p_1,...,p_n)[q]= \sum
\limits_{\vec{m}}
 U_{{(\Gamma/\gamma)}^{\vec{m}}}(p_1,...,p_n) \bigotimes
 \limits_{\Gamma/\gamma} \delta (\alpha)[q].
\end{equation}
This quantity depends on loop momenta of \(\Gamma/\gamma\). Here
we use denotations from the page (11). It is easy to see, (as in
lemma 2), that \(U^{z}_{\Gamma}(p)(\vec{\alpha})[q]\) can be
multiplied by \(\delta(\lambda-|\vec{\alpha}|)\) and it is
possible to extract integration on \(\lambda\) (If \rm Re \mit z
is enough large).\newline

 Let \(\Gamma\) be one particle irreducible diagram and
 \(\gamma_1,...,\gamma_k \) be a set of one particle
 irreducible subdiagrams such that for each \(i, j=1,...,k\;i\neq j\)
 the sets of vertices of diagrams \(\gamma_i\) and \(\gamma_j\) do
 not intersect.
 Let us define the quotient diagram
\begin{equation}
\Gamma/{\gamma_1\star....\star\gamma_k }.
\end{equation}
 We replace each subdiagram \(\gamma_i\) by a vertex \(\tilde{v}_i\)
 and put
\begin{equation}
\varphi_{\Gamma/{\gamma_1\star....\star\gamma_k
}}(v)=\varphi_\Gamma(v)
\end{equation}
if \(v\neq\tilde{v}_i\) and \(\forall i=1,...,k\)
\begin{equation}
\varphi_{\Gamma/{\gamma_1\star....\star\gamma_k }}(\tilde{v}_i)=1.
\end{equation}
Let \(\gamma\subset\gamma'\subset\Gamma\) be one particle
irreducible diagrams. Let \([q]_{\Gamma/\gamma'}\) be a set of
loop momenta of \(\Gamma/\gamma'\). Let us define \((C_\gamma
\star U_\Gamma)(\vec{\alpha})(p_1,...,p_n)[q]_{\Gamma/\gamma'}\)
by the formula
\begin{eqnarray}
(C_\gamma \star
U_\Gamma)(\vec{\alpha})(p_1,...,p_n)[q]_{\Gamma/\gamma'}\nonumber\\
=\sum \limits_{\vec{m}} \int [dq]_{\gamma'/\gamma}
 (U_{(\Gamma/\gamma)}^{\vec{m}}(p_1,...,p_n)[q]_{\Gamma/\gamma} \bigotimes
 \limits_{\Gamma/\gamma} \delta^{(\vec{m})}) (\vec{\alpha}).
 \end{eqnarray}
 Now let us consider the following situation. We have two one particle irreducible
 diagrams \(\gamma\subset\gamma'\subset\Gamma\) and a set \(A\) of painted lines of
 \(\Gamma\) which are satisfy the following condition:
 If we replace \(\gamma'\) by a point the set \(A\) becomes a tree.
 Let \(\tilde{\eta}_A\) be a function which have been previously described.
 In this situation the following lemma holds. \newline

 \textbf{Lemma.} If \(\rm Re \mit\; z\) is enough large than:
 \begin{eqnarray}
 \langle (C_\gamma \star
 U_\Gamma)(\vec{\alpha})(p)\delta(|\vec{\alpha}|-\lambda),
 \tilde{\eta}_A (\vec{\alpha})g(\vec{\alpha})\rangle\nonumber\\
 =\sum \limits_{\vec{m}} \int [dq]_{\Gamma/\gamma'}
 \langle ( U_{(\Gamma/\gamma)}^{\vec{m}}(p)[q]_{\Gamma/\gamma'} \bigotimes
 \limits_{\Gamma/\gamma} \delta^{(\vec{m})}) (\vec{\alpha})\delta(|\vec{\alpha}|-\lambda),
 \tilde{\eta}_A(\vec{\alpha})g(\vec{\alpha})\rangle. \label{ZU}
 \end{eqnarray}
\textbf{Proof.}\newline
 Let us divide variables \(\alpha\) into three groups:\newline
 By definition variables \(\beta'\) corresponds to the lines of \(\gamma\).\newline
 Variables \(\beta''\) corresponds to the lines of
 \(\gamma'/\gamma\).\newline
 Variables \(\beta'''\) corresponds to  the lines of \(\Gamma/\gamma'\). We have \newline
\begin{eqnarray}
\langle (C_\gamma \star
 U_\Gamma)(\vec{\alpha})(p)\delta(|\vec{\alpha}|-\lambda),
 \tilde{\eta}_A (\vec{\alpha})g(\vec{\alpha})\rangle \nonumber\\
=\sum \limits_{\vec{m}} \langle \delta^{(\vec{m})}(\vec{\beta}')
\langle
U_{(\Gamma/\gamma)}^{\vec{m}}(p)(\vec{\beta}'',\vec{\beta}''')
 \delta(|\vec{\beta}'|+|\vec{\beta}''|+|\vec{\beta}'''|-\lambda),\nonumber\\
 \tilde{\eta}_A(\vec{\beta}',\vec{\beta}'',\vec{\beta}''')
 g(\vec{\beta}',\vec{\beta}'',\vec{\beta}''')\rangle.
\end{eqnarray}
We have
\begin{eqnarray}
\langle
U_{(\Gamma/\gamma)}^{\vec{m}}(p)(\vec{\beta}'',\vec{\beta}''')
 \delta(|\vec{\beta}'|+|\vec{\beta}''|+|\vec{\beta}'''|-\lambda),\nonumber\\
 \tilde{\eta}_A(\vec{\beta}',\vec{\beta}'',\vec{\beta}''')
 g(\vec{\beta}',\vec{\beta}'',\vec{\beta}''')\rangle\nonumber\\
 =\int d\vec{\beta}''d\vec{\beta}'''
 \delta(|\vec{\beta}'|+|\vec{\beta}''|+|\vec{\beta}'''|-\lambda),\nonumber\\
=\{ \tilde{\eta}_A(\vec{\beta}',\vec{\beta}'',\vec{\beta}''')
 g(\vec{\beta}',\vec{\beta}'',\vec{\beta}''')
 \int [dq]_{\Gamma/\gamma'} U_{(\Gamma/\gamma)}^{\vec{m}}(p) [q]_{\Gamma/\gamma'}
 (\vec{\beta}'',\vec{\beta}''')\} \label{58}.
\end{eqnarray}
It is easy to see that all variables \(\vec{\beta}'''\) except
variables corresponding to some tree graph separated from zero.
Variables \(\vec{\beta}''\) take values in some bounded region.
The function \( U_{(\Gamma/\gamma)}^{\vec{m}}(p)
[q]_{\Gamma/\gamma'}\) is a function of fast decay at infinity (as
\(e^{-A|q|^2}\), \(A>0\)) on \([q]_{\Gamma/\gamma'}\). Moreover,
if
 \([q]_{\Gamma/\gamma'}\) and \(\vec{\beta}',\vec{\beta}''\) take a
 value in some bounded region, this function is uniformly continuous on
 \([q]_{\Gamma/\gamma'}\) and \(\vec{\beta}',\vec{\beta}''\) in this region.
 By using this remark we can approximate the integral over \(d [q]_{\Gamma/\gamma'}\) in (\ref{58})
 by Riemann sum. By other words:
\begin{eqnarray}
\sum \limits_{[q_i]_{\Gamma/\gamma'} } [\Delta q]_{\Gamma/\gamma'}
\int d\vec{\beta}''d\vec{\beta}'''
 \delta(|\vec{\beta}'|+|\vec{\beta}''|+|\vec{\beta}'''|-\lambda),\nonumber\\
\{ \tilde{\eta}_A(\vec{\beta}',\vec{\beta}'',\vec{\beta}''')
 g(\vec{\beta}',\vec{\beta}'',\vec{\beta}''')
  U_{(\Gamma/\gamma)}^{\vec{m}}(p) [q_i]_{\Gamma/\gamma'}
 (\vec{\beta}'',\vec{\beta}''')\}\longrightarrow {59}.\label{60}
\end{eqnarray}
uniformly on \(\beta'\) at each compact \(\vec{\beta}'\).\newline
 Analogously, it is easy to prove, that the left hand side of
 (\ref{60}) tends to the right hand side of (\ref{60}) in the
 topology of \(C^{|\vec{m}|}\) at each compact if \(\rm Re \mit\; z\) is enough large. Therefore the
 Riemann sums for the integral in the right hand side of
 (\ref{ZU}) tend to the left hand side of (\ref{ZU}).
 It is easy to prove that the integral in the right hand side exist also in the Lebegues sense.
 The lemma is proved.

 The trivial generalization of the previous lemma is the following lemma.  \newline

\textbf{Lemma.} Let \(A\) be a some subset of the set of lines of
\(\Gamma\), and \(\gamma_1^A,...,\gamma_f^A\) be corresponding
subdiagrams. Let us divide the parameters  \(\{\alpha\}\) into
\(f+1\) groups; \(\vec{\beta}'_i\) are parameters corresponding to
\(\gamma_i^A\) and \(\vec{\beta}''\) are the other parameters. If
\(\rm Re \mit\; z\) is enough large, we have:
\begin{eqnarray}
 \langle\tilde{R}^z_\Gamma(\vec{\alpha})(p)\delta(1-|\vec{\alpha}|),
\tilde{\eta}_A(\vec{\alpha})g(\vec{\alpha})\rangle\nonumber\\
 =\int
[dq]_{\Gamma/{\gamma_1^A\star...\star\gamma_f^A}}\nonumber\\
\langle \bigotimes \limits_{i=1}^{f} R^z_{\gamma_i}(\vec{\beta}'_i)(p)
[q]_{\Gamma/{\gamma_1^A\star...\star\gamma_f^A}} \{\int d
\vec{\beta}'' \delta(1-\sum \limits_{i=1}^f
|\vec{\beta}'|-|\vec{\beta}''|)\nonumber\\
\tilde{\eta}_A( \vec{\beta}'_1,...,\vec{\beta}'_f,\vec{\beta}'')
 U_{\Gamma/{\gamma_1^A\star...\star\gamma_f^A}}(\vec{\alpha})
 [q]_{\Gamma/{\gamma_1^A\star...\star\gamma_f^A}}(p)
 g(\vec{\beta}'_1,...,\vec{\beta}'_f,\vec{\beta}'')\}\rangle.
 \label{Zu}
 \end{eqnarray}
 Here \(R^z_\Gamma(\vec{\beta}'_i)(p)
[q]_{\Gamma/{\gamma_1^A\star...\star\gamma_f^A}} \) depends on the
external momenta and the loop momenta
\([q]_{\Gamma/{\gamma_1^A\star...\star\gamma_f^A}} \). We will
prove below that the right hand side of ( \ref{Zu}) has an unique
analytical on whole open complex plane except \(D \setminus
\{0\}\) (\(D\) is some discrete set). Analogously:
\begin{eqnarray}
\langle
\tilde{R}^z_\Gamma(\vec{\alpha})(\frac{p}{\sqrt{\lambda}})\delta(\lambda-|\vec{\alpha}|),
\tilde{\eta}_A(\frac{\vec{\alpha}}{\lambda})g(\vec{\alpha})\rangle\nonumber\\
 =\int [dq]_{\Gamma/{\gamma_1^A\star...\star\gamma_f^A}}
\lambda^{\sharp R_\Gamma-1+z\sharp
R_{{\Gamma/{\gamma_1^A\star...\star\gamma_f^A}}}-\frac{1}{2}\{4
L_{{\Gamma/{\gamma_1^A\star...\star\gamma_f^A}}}+\sum \limits_{v
\in {\Gamma\setminus(\gamma_1^A\cup...\cup\gamma_f^A)}} \rm deg \mit
\varphi(v)\}}
\nonumber\\
\int d \vec{\beta}'' \langle \bigotimes \limits_{i=1}^{f}
(\Lambda_{\gamma_i} R^z_{\gamma_i}) (\vec{\beta}'_i)(p)
[q]_{\Gamma/{\gamma_1^A\star...\star\gamma_f^A}} \{ \delta(1-\sum
\limits_{i=1}^f
|\vec{\beta}'|-|\vec{\beta}''|)\nonumber\\
\tilde{\eta}_A( \vec{\beta}'_1,...,\vec{\beta}'_f,\vec{\beta}'')
 U_{\Gamma/{\gamma_1^A\star...\star\gamma_f^A}}(\vec{\alpha})
 [q]_{\Gamma/{\gamma_1^A\star...\star\gamma_f^A}}(p)
 g(\lambda \vec{\alpha})\}\rangle. \label{ZZ}
 \end{eqnarray}
 Here \(L_{{\Gamma/{\gamma_1^A\star...\star\gamma_f^A}}}\)
 is a number of loop of the diagram.

 \section{Estimates of the integrand in the integral over
 d\(\lambda\)}
 Note that the expression in the figured bracket in (\ref{ZZ}) is an
infinitely differentiable
 function and its norm \(\|\|_m\) \(\forall m\) admit an estimate:
 there exists a constant \(A>0\) such that for each \(N\) there exists a constant \(C>0\)
 and a polynomial \(P(p,[q]_{\Gamma/{\gamma_1^A\star...\star\gamma_f^A}})\)
 such that
\begin{eqnarray}
\|\cdot\|_m\leq
C|P(p,[q]_{\Gamma/{\gamma_1^A\star...\star\gamma_f^A}})|
\frac{1}{(1+\lambda)^N}e^{-A[q]_{\Gamma/{\gamma_1^A\star...\star\gamma_f^A}}^2},
\end{eqnarray}
for some polynomial \(P\) and an arbitrary positive integer \(m\).
The constant \(C\) depends on \(N\). We have by using the
inductive assumption d) ii)
\begin{eqnarray}
|\langle
\tilde{R}^z_\Gamma(\vec{\alpha})(\frac{p}{\sqrt{\lambda}})\delta(\lambda-|\vec{\alpha}|),
\tilde{\eta}_A(\frac{\vec{\alpha}}{\lambda})g(\vec{\alpha})\rangle|\nonumber\\
\leq C \frac{\lambda^{-\varepsilon}}
{(1+\lambda)^N}\Pi\{\lambda^{-\frac{\Omega_{\gamma_i^A}}{2}}\nonumber\\
\lambda^{(1+z)\sharp
R_{{\Gamma/{\gamma_1^A\star...\star\gamma_f^A}}}-1-\frac{1}{2}\{4
L_{{\Gamma/{\gamma_1^A\star...\star\gamma_f^A}}}+\sum \limits_{v
\in {\Gamma\setminus(\gamma_1^A\cup...\cup\gamma_f^A)}}
deg\varphi(v)\}}\}\nonumber\\
\leq C \frac{\lambda^{-\varepsilon}} {(1+\lambda)^N}
\lambda^{-\frac{\Omega_\Gamma}{2}-1}\lambda^{-x_\Gamma (\rm Re \mit\;
z)^{-}}.
\end{eqnarray}

\textbf{Estimates on derivatives on external momenta.} By using
differentiability by external momenta and inductive assumption on
\(R\) we have:

\begin{eqnarray}
|\langle
{}^{(l)}\tilde{R}^z_\Gamma(\vec{\alpha})(p)\delta(\lambda-|\vec{\alpha}|),
\tilde{\eta}_A(\frac{\vec{\alpha}}{\lambda})g(\vec{\alpha})\rangle|\nonumber\\
\leq C \frac{\lambda^{-\varepsilon+\frac{l}{2}}} {(1+\lambda)^N}
\lambda^{-\frac{\Omega_\Gamma}{2}-1}\lambda^{-x_\Gamma (\rm Re \mit\;
z)^{-}}\label{65}.
\end{eqnarray}

\textbf{Holomorphic property.} Let us prove \begin{equation}
\langle
\tilde{R}^z_\Gamma(\vec{\alpha})(\frac{p}{\sqrt{\lambda}})\delta(\lambda-|\vec{\alpha}|),
\tilde{\eta}_A(\frac{\vec{\alpha}}{\lambda})g(\vec{\alpha})\rangle
\end{equation}
is a holomorphic function of \(z\) on the \(\{z \in
\mathbb{C}|\rm Im \mit\;z>A\}\) for some positive \(A\). This quantity can be
represented as an integral over
\([dq]_{\Gamma/{\gamma_1^A\star...\star\gamma_f^A}}\). It is clear
that the integrand holomorphic on \(z\). By using the Cauchy
estimates, we find that the derivative on \(z\) of the integrand
increases slowly than some polynomials \(\times e^{-Ap^2}\). From
other hand the derivative of the integrand on the loop momenta
increase analogously. Therefore the integrand is uniformly
continuous on \(z\) and \(p\) if \(z\) belongs to some compact
\(K\) such that \(K \subset \{z|z \in \mathbb{C}\;:\rm Im \mit\; z
>A\} \) for some positive \(A\). Therefore the integrand is measurable. Analogously one can
prove that (\ref{65}) is continuous on \(z\) on just described
set. Let \(\pi\) be a small contour homotopic in \(\{z \in
\mathbb{C}|z>A\}\). Let us consider the integral of absolute value
of integrand in (\ref{ZZ}) over the direct product of our contour
and the space of external momenta. If we integrate at first
(\ref{ZZ}) over \(\pi\) by \(z\) and at second by loop momenta we
obtain a zero. By using Morera's and Fubini's theorem one has that
(\ref{65}) is holomorphic on \(z\) in our region.\newline
\section{End of the proof. Check of the inductive assumptions}
 \textbf{The analytical extension of \(\tilde{R}^z_\Gamma(\vec{\alpha})(p)\).}
 Let \(M_{\Omega_\Gamma}\) be a Taylor projector which to each function
 of external momenta of diagram \(\Gamma\) assigns its Taylor polynomial
  with the center at zero of degree \(\Omega_\Gamma\).
 \(1-M_{\Omega_\Gamma}\) by Schlomilch theorem can be expressed through the integral
 of partial derivatives of degree \(\Omega_\Gamma+1\). By using the estimate (\ref{65})
 and an argumentation based on The Morera theorem we find, that \((1-M_{\Omega_\Gamma})\tilde{R}^z\)
 has an analytical continuation into some neighborhood of zero. If
 we use the substitution
  \(\vec{\alpha}\mapsto\lambda\vec{\alpha}\),
 \(p\mapsto\frac{p}{\sqrt{\lambda}}\) and estimate (64) we find,
 that
\begin{equation}
 {}^{(l)}(1-M_{\Omega_\Gamma})\tilde{R}^z \in \Upsilon_{\Gamma}^{m,((1/2)
\Omega_{\Gamma} + \sharp R_{\Gamma}-\frac{|l|}{2} + x^{l}_{\Gamma}
{(\rm Re \mit\; z)}^{-})} \label{67}.
\end{equation}
In other words for \((1-M_{\Omega_\Gamma})\tilde{R}^z\) the
inductive assumption d) ii) holds.

Now let us investigate
 \(M_{\Omega_\Gamma}\tilde{R}^z(\vec{\alpha})(p)\). Let \(M^n_\Gamma\) be a
projector on  homogenous polynomials of degree \(n\). We have
 \begin{eqnarray}
 M^n_\Gamma \tilde{R}^z(\vec{\alpha})(p)\nonumber\\
 =
 \int \limits_{0}^{\infty} d\lambda \lambda^{\sharp R-1+\frac{n}{2}} \langle
\Lambda_\lambda(M^n_\Gamma\tilde{R}^z_\Gamma)(\vec{\alpha})(p)
\delta(1-|\vec{\alpha}|),g(\lambda\vec{\alpha})\rangle. \label{68}
\end{eqnarray}
For each diagram \(\Gamma\) \(\tilde{R}^z_\Gamma\) can be
decomposed (by inductive assumption) into the sum of homogenous
functions with respect the operation \(\Lambda_\lambda\)
\({(R^z_\Gamma)}^{\delta}\) of degree
\(\frac{\Omega_\gamma}{2}+\sharp R_\gamma-y^\delta_\gamma z\).
Therefore (\ref{67}) is the sum of terms of degree
\begin{eqnarray}
 M_\Gamma^n\tilde{R}^z(\vec{\alpha})(p)\nonumber\\
 =\sum \limits_A \int \limits_{0}^{\infty} d\lambda \lambda^{-1
-\frac{\Omega_\Gamma}{2}+y^\delta_{\Gamma,A} z+\frac{n}{2}}
\langle M^n_\Gamma(\tilde{R}^z_\Gamma)^\delta(\vec{\alpha})(p)
\delta(1-|\vec{\alpha}|),\tilde{\eta}_A
g(\lambda\vec{\alpha})\rangle
\end{eqnarray}
for some \(y^\delta_{\Gamma,A}\). By using the standard analytical
property of the distribution \(\theta(\lambda)\lambda^z\) we find,
that (\ref{68}) has an analytical continuation into some
punctured neighborhood of zero. So we have prove the inductive
assumption b).\newline

 \textbf{Locality of counterterms \(C_\Gamma\)(beginning).}
  \(C_\Gamma\) is a polynomial on the external momenta. This fact follows from
  the fact that \((1-M_{\Omega_\Gamma})\tilde{R}^z\) has an analytical continuation
  into some neighborhood of zero.
 Let us prove that \(C_\Gamma\) is a finite linear combination of \(\delta\)-functions on \(\vec{\alpha}\).
 The pole part came from \(M^n_\Gamma \tilde{R}^z_\Gamma\), more
 precisely from:
\begin{eqnarray}
 \int \limits_{0}^{1} d\lambda \lambda^{\sharp R-1+\frac{n}{2}} \langle
\Lambda_\lambda(M^n_\Gamma\tilde{R}^z_\Gamma)(\vec{\alpha})(p)
\delta(1-|\vec{\alpha}|),g(\lambda\vec{\alpha})\rangle. \label{69}
\end{eqnarray}
Let us write a representation
\begin{eqnarray}
g(\lambda\vec{\alpha})=\sum \limits_{k=1}^{\Phi}
\frac{\lambda^k}{k!}(\frac{d}{d\lambda})^k g(\lambda
\vec{\alpha})|_{\lambda=0}+\lambda^{\Phi+1}\psi(\lambda)
\end{eqnarray}
\(\psi(\lambda)\) is a smooth function. The contribution into the
pole part comes from the terms of the form
\begin{eqnarray}
 \int \limits_{0}^{1} d\lambda \lambda^{\sharp R-1+\frac{n}{2}} \langle
\Lambda_\lambda(\tilde{R}^z_\Gamma)(\vec{\alpha})(p)
\delta(1-|\vec{\alpha}|),\tilde{\eta}_A (\frac{d}{d\lambda})^k
g(\lambda \vec{\alpha})|_{\lambda=0})\rangle.
\end{eqnarray}
These terms has an analytical continuation into some punctured
neighborhood of the point \(z=0\). We have to prove that these
terms have the form of linear combination of  \(\delta\)-functions
and their derivatives. But
\begin{eqnarray}
g(0\vec{\alpha})=g(0),\nonumber\\
 \frac{d}{d\lambda}g(\lambda\vec{\alpha})=\sum \limits_i \alpha_i
 (\frac{\partial}{\partial\alpha_i}g(\vec{\alpha})|_{\vec{\alpha}=0}),
\end{eqnarray}
e.c.t.

 The proof of the point d) i) is analogues to the proof of
 the point b). We need only the Taylor projector \(M_{\Omega_\Gamma}\) replace by \(M_N\)
 and chose the number \(N\) enough large.

 \textbf{The proof of the points d) iii) and d) iv).} We have proved
 before that the functions \(\tilde{R}^z\) is a sum of meromorphic functions with respect to the
 operation \(\Lambda_\lambda\).  This
 fact and the inductive assumptions imply that the function \(U_\Gamma\) is meromorphic too.
 This fact and the formula for \(R\)-operation implies that the function \(\tilde{R}^z\)
 can be represented as a sum of homogenous function with proper coefficients of homogeneity.

 \textbf{Locality of counterterms \(C_\Gamma\) (the end).} Now let
 us prove that the constructed counterterms have a proper power of
 homogeneity with respect \(\Lambda_\lambda\).
 This statement follows from the facts that counterterms are \newline
 a) the linear combination of homogenous functions,\newline
 b) the sum of the pole parts of holomorphic functions in some
 neighborhood of zero, which power of homogeneity is equal to
 \(\frac{\Omega_\Gamma}{2}+\sharp R_\gamma-y^\delta_\Gamma z\).
 So the inductive assumption c) is proved.

 The inductive assumption  a) follows from the form of subtract operator.

\textbf{Proof of the point d) ii).} The fact that
\begin{equation}
 {}^{(l)}(1-M_{\Omega_\Gamma})\tilde{R}^z_\Gamma \in \Upsilon_{\Gamma}^{m,((1/2)
\Omega_{\Gamma} + \sharp R_{\Gamma}-\frac{l}{2} + x^{l}_{\Gamma}
{(\rm Re \mit\; z)}^{-})}
\end{equation}
if \(\rm Re \mit\; z \geq -\epsilon_\Gamma\) for some
\(\epsilon_\Gamma\) is proved. \(M_{\Omega_\Gamma}\tilde{R}^z_\Gamma\)
can be represented as a linear combination of homogenous functions
with proper power of homogeneity, so  \(\tilde{R}^z_\Gamma \in
\Upsilon_{\Gamma}^{m,((1/2) \Omega_{\Gamma} + \sharp
R_{\Gamma}-\frac{l}{2} + x^{l}_{\Gamma} {(\rm Re \mit\; z)}^{-})}\) if
\(\rm Re \mit z \geq -\epsilon_\Gamma\) except probably zero.
 \(C_\Gamma^z\) is homogenous and its homogenous power is equal to \(\Omega_\Gamma+\sharp
 R\).
\(\tilde{R}^z_\Gamma\) belongs to the needed class if \(\rm Re \mit \;z \geq
-\epsilon_\Gamma\) except probably zero. The fact that
\(\tilde{R}^z_\Gamma\) belongs to the needed class if \(\rm Re
\mit\; z \geq -\epsilon_\Gamma\) can be proven by using the Cauchy
theorem.

\section{Conclusion.} In the present paper we gave some new proof of the Bogoliubov ---
Parasiuk theorem based on the theory of distributions. This technique will be used in the next
paper of this series to prove that the divergences in nonequilibrium diagram technique can be
renormalized by the counterterms of the asymptotical state.

Author is grateful to Yu. E. Lozovik and I.L. Kurbakov for very useful discussions.

\end{document}